\documentclass[twoside,a4paper,11pt]{sea22}
\usepackage{graphicx}
\usepackage{hyperref}
\usepackage{media9}
\begin{document}
\pagenumbering{arabic}
\pagestyle{myheadings}
\thispagestyle{empty}
{\flushleft\includegraphics[width=\textwidth,bb=58 650 590 680]{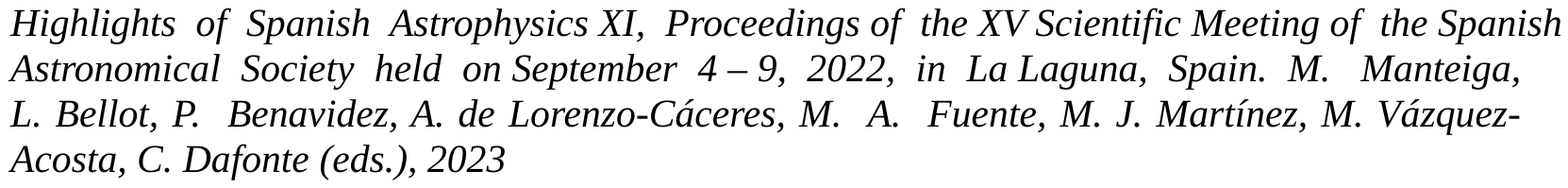}}
\vspace*{0.2cm}
\begin{flushleft}
{\bf {\LARGE
%
Post-red-giant-branch Planetary Nebulae
%
}\\
\vspace*{1cm}
%
Jones, D.$^{1,2,3}$,
Hillwig, T.C.$^4$,
and Reindl, N.$^5$
%
}\\
\vspace*{0.5cm}
%
$^{1}$
Instituto de Astrof\'isica de Canarias, E-38205 La Laguna, Tenerife, Spain\\
$^{2}$
Departamento de Astrof\'isica, Universidad de La Laguna, E-38206 La Laguna, Tenerife, Spain\\
$^{3}$
Nordic Optical Telescope, Rambla Jos\'e Ana Fern\'andez P\'erez 7, 38711, Bre\~na Baja, Spain\\
$^4$
Department of Physics and Astronomy, Valparaiso University, Valparaiso, IN 46383, USA\\
$^5$
Institute for Physics and Astronomy, University of Potsdam, Karl-Liebknecht-Str. 24/25, D-14476 Potsdam, Germany
%
\end{flushleft}
%
\markboth{
Post-RGB PNe
}{ 
%
David Jones
%
}
\thispagestyle{empty}
\vspace*{0.4cm}
\begin{minipage}[l]{0.09\textwidth}
\ 
\end{minipage}
\begin{minipage}[r]{0.9\textwidth}
\vspace{1cm}
\section*{Abstract}{\small
Common envelope events have been associated with the formation of a planetary nebulae since its proposition more than forty five years ago.  However, until recently there have been doubts as to whether a common envelope while the donor is ascending the red giant branch, rather than the subsequent asymptotic red giant branch, would result in a planetary nebula.  There is now strong theoretical and observational evidence to suggest that some planetary nebulae are, indeed, the products of common envelope phases which occurred while the nebular progenitor was on the red giant branch.  The characterisation of these systems is challenging but has the potential to reveal much about the common envelope -- a critical evolutionary phase in the formation of a plethora of interesting astrophysical phenomena.
%
\normalsize}
\end{minipage}
%
%
%
\section{Introduction \label{intro}}
The common envelope (CE) phase is perhaps the most poorly understood aspect of close-binary evolution \cite{ivanova13}, in spite of its critical nature for the formation of a number of important phenomena -- ranging from type \textsc{i}a supernovae (irrespective of their formation channel) through to merging binary black holes recently observed in gravitational waves.  A CE is formed following run-away unstable mass transfer in a close binary and ultimately leads to dramatic in-spiral of the binary, sometimes to merger, and the ejection of the donor's envelope (partial ejection in the case of a merger).  Planetary nebulae (PNe) have long been seen as important probes of the CE, being identified by Paczy\'nski (one of the first to outline the CE scenario) as the critical proof of its existence  \cite{paczynski76}.  They are of particular interest because post-CE PNe are the immediate after-products of the CE (where the central stars have yet to have time to relax/evolve), and the only systems where the ejected envelope is still observable\footnote{The envelopes surrounding (Luminous) Red Novae represent the partial ejection of the evolution as the stars merged prior to complete ejection \cite{kaminski21}.}  (as the PN itself).

Until recently, it was thought that observable PNe would only be formed following CE events occurring when the donor was on the asymptotic giant branch (AGB), or close to the very tip of the red giant branch (RGB).  Otherwise, the post-RGB post-CE remnants were expected to evolve too slowly or to too low a maximum temperature in order to ionise the ejected material before it dissipated into the surrounding interstellar medium \cite{iben93}.  Updated simulations by Hall et al.\ have since shown that most CE events on the RGB should, in fact, lead to observable PNe, revising the minimum mass limit for the post-RGB central star down to around 0.25~M$_\odot$ (c.f.\ the previous estimate of $>$0.4~M$_\odot$).  This is particularly interesting in understanding the population of PNe as a whole given that at least 20\% of all PNe are found to be the products of a CE phase \cite{jones17} and the majority of post-CE binaries are post-RGB \cite{rebassa-mansergas11}. As such, a number of recent studies have focussed on the search for and characterisation of post-RGB central stars of PNe, and their implications for our understanding of the CE.  In this proceedings, I present a review of this recent work and its future prospects.

\section{Candidates \label{candidates}}	

Roughly eight strong candidate post-RGB central stars have thus far been identified (see table \ref{tab1}).  Although the central stars are orders of magnitude more luminous than their low-mass (M-type) companions, these systems are generally double-lined to the high levels of irradiation leading to the formation of spectroscopically detectable emission lines on the day face of the companions.  Similarly, even when they are not eclipsing, the central star systems present strong photometric variability due to the differing projection of the irradiated face of the companion as a function of orbital phase.  Here, we will discuss further the properties of each of these candidates as determined by detailed modelling of light and radial velocity curves, and/or through non-local thermal equilibrium (NLTE) modelling of the spectrum of the hot post-RGB component.

\begin{table}[h] 
\caption{Candidate post-RGB central stars of PNe} 
\center
\begin{minipage}{0.5\textwidth}
\center
\begin{tabular}{lcccl} 
\hline\hline 
PN & RA  & Dec & Period & Reference\\ [0.5ex]   
\hline 
Abell~46 & 18:31:18.3 & $+$26:56:12.9 & 0.472~d &\cite{afsar08}\\
ESO~330-9 & 16:02:18.1 & $-$41:26:49.5 & 0.296~d & \cite{hillwig17}\\
HaTr~4 & 16:45:00.2 & $-$51:12:21.1 & 1.738~d & \cite{hillwig16}\\
HaTr~7 & 17:54:09.5 & $-$60:49:57.7 & 0.322~d & \cite{hillwig17}\\
HaWe~13 & 19:31:07.2 & $-$03:42:31.5 & -- & \cite{frew16,napiwotzki99}\\
Hf~2-2 & 18:32:30.8 & $-$28:43:20.0 & 0.399~d & \cite{hillwig16}\\
Ou~5 & 21:14:20.0 & $+$43:41:36.0 & 0.364~d & \cite{jones22}\\
PN~G283.7$-$05.1 & 09:58:32.3 & $-$61:26:40.0 & 0.246~d & \cite{jones20}\\ [1ex]  
\hline
\end{tabular} 
\end{minipage}
\label{tab1} 
\end{table}

\subsection{Abell 46}

\begin{figure}[!h]
\center
\includegraphics[width=9cm,angle=0,clip=true]{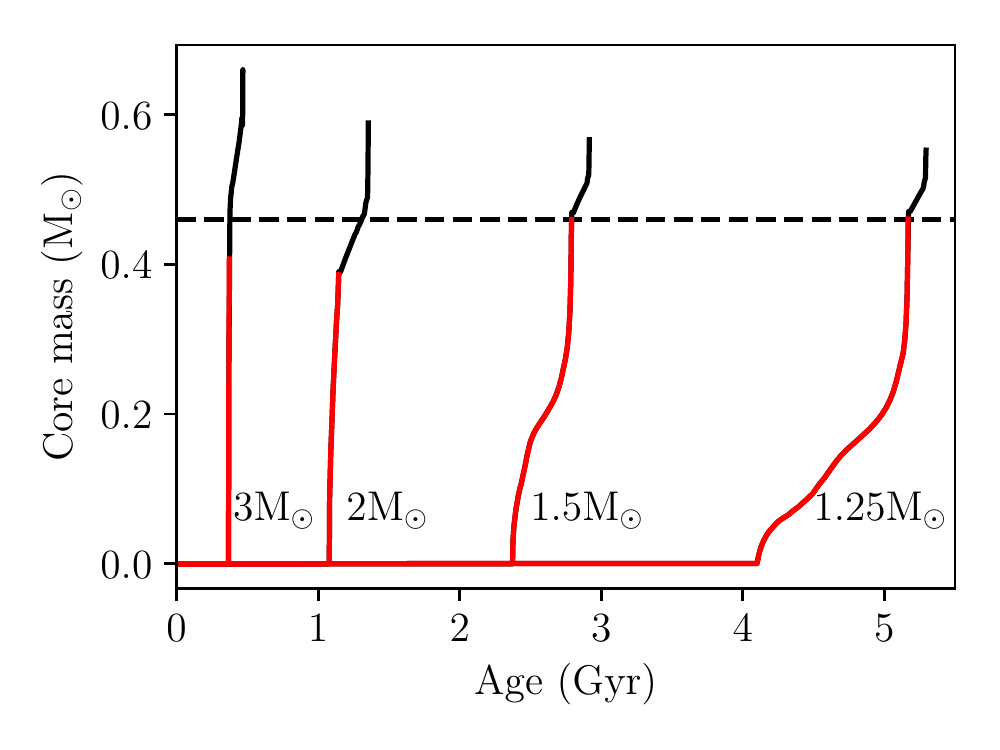} 
\caption{\label{fig1} The core mass evolution of stars for a range of initial masses. The evolution to the tip of the RGB is shown in red with the subsequent AGB evolution in black.  This demonstrates that the maximum mass for the post-RGB central star is approximately 0.46~M$_\odot$.}
\end{figure}

Simultaneous light and radial velocity curve modelling of the HW Vir-type central star of Abell~46 derives a mass of $0.51\pm0.05$~M$_\odot$ with a luminosity of $\log L\sim2.2$ \cite{afsar08}.  While the derived mass is a little on the high side for a post-RGB central star, the uncertainties do clearly encompass post-RGB masses (see figure \ref{fig1}).  The low luminosity on the other hand is entirely consistent with a post-RGB remnant.  Furthermore, the nebula has been found to have a very large abundance discrepancy factor, a property speculatively linked with post-RGB evolution \cite{jones16}.  Interestingly, the surface gravity and temperature implied by the light and radial velocity curve 
modelling are consistent with a post-RGB evolutionary track for a much lower mass than that derived by the modelling (see figure \ref{fig2}).  An NLTE modelling of the central star spectrum could better constrain the effective temperature and surface gravity, placing stronger constraints on the possible post-RGB nature of Abell~46.

\subsection{ESO~330-9}

NLTE modelling of the spectrum of the central star of ESO~330-9 gives a surface gravity and effective temperature consistent with post-RGB evolutionary tracks (see figure \ref{fig2}).  Furthermore, an excellent fit to the light and radial velocity curves of the binary could be found constraining the parameters of the primary to match those derived from the NLTE modelling \cite{hillwig17}.  The light curve itself is not eclipsing so this is perhaps not a stringent test of the system parameters but, nonetheless, ESO~330-9 is an excellent candidate post-RGB central star.

\subsection{HaTr~4}

Multi-band modelling of the eclipsing light curve of HaTr~4 results in a range of radii and temperatures consistent with post-RGB evolution \cite{hillwig16}.  Unfortunately, without radial velocities to properly constrain the mass or NLTE spectral modelling, it is difficult to assess the true nature of the central star. 
\subsection{HaTr~7}

\begin{figure}[!h]
\center
\includegraphics[width=11cm,angle=0,trim= 0 120 0 120, clip=true]{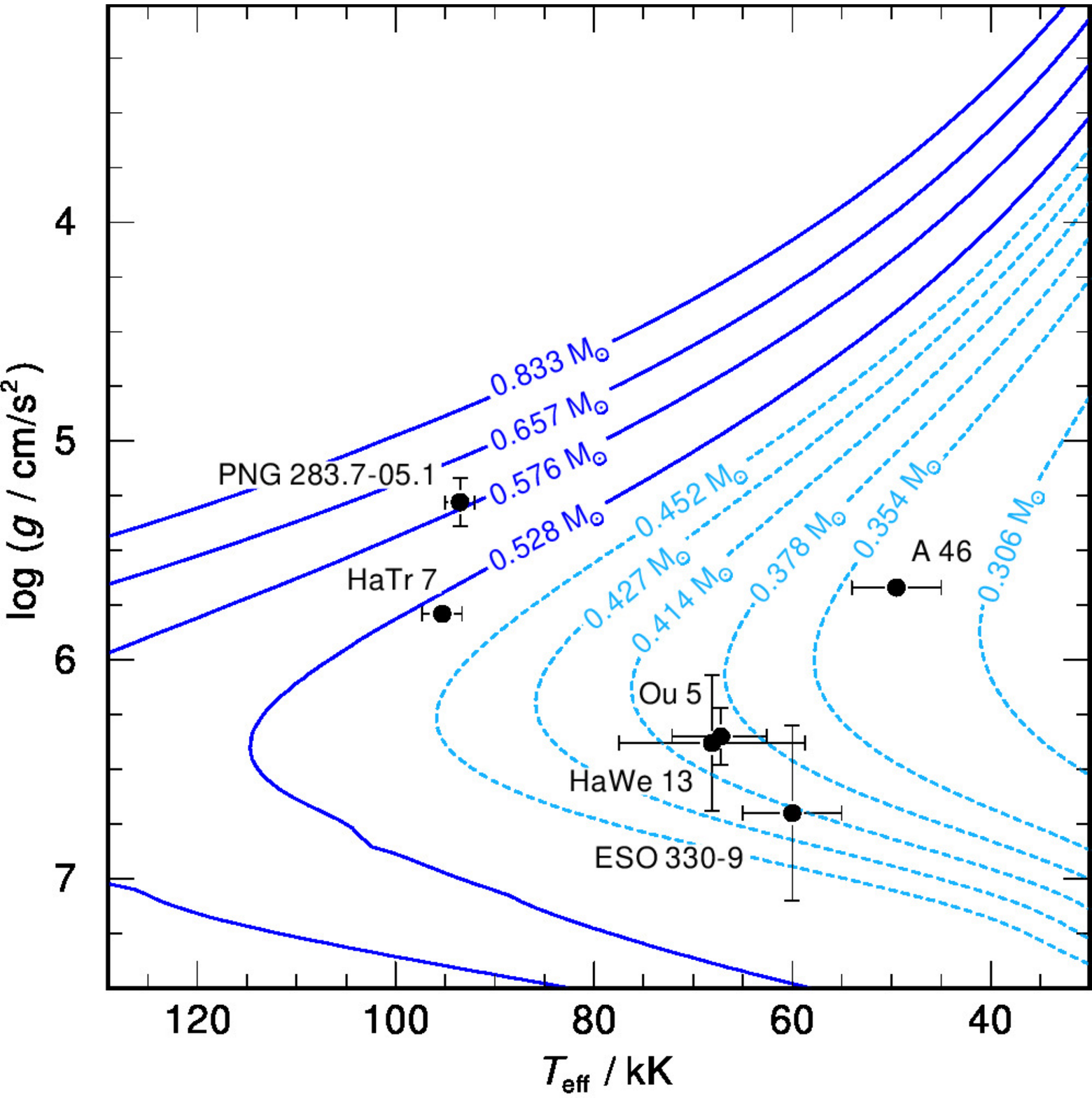} 
\caption{\label{fig2} A Kiel diagram showing the properties of the five strongest candidate post-RGB stars for which effective temperatures and surface gravities have been measured (either spectroscopically or via combined light and radial velocity curve modelling.  Figure reproduced from \cite{jones22}.}
\end{figure}

NLTE modelling of the spectrum of the central star of HaTr~7 gives parameters which lie between post-RGB and post-AGB tracks on the Kiel diagram (figure \ref{fig2}).  Just as for ESO~330-9, an excellent fit to the non-eclipsing light curve and radial velocity curves could be found with parameters which match those predicted by the NLTE modelling \cite{hillwig17}.  

\subsection{HaWe~13}

The central star of HaWe~13 was identified as being under-luminous based on its gravity distance, while the nebula presents a morphology consistent with being the product of a CE phase \cite{frew16}.  However, no study has as yet definitively revealed the binary nature of the central star.

\subsection{Hf~2-2}

Just as for HaTr~4, modelling of multi-band non-eclipsing light curves indicates a low luminosity central star. A single spectrum of the central star was also taken allowing the mass ratio to also be constrained, indicating a low-mass primary \cite{hillwig16}.

\subsection{Ou~5}

Simultaneous light and radial velocity curve modelling of the eclipsing binary (HW Vir-type) at the heart of Ou~5 derives a mass of $0.50\pm0.06$~M$_\odot$ with a luminosity of $\log L\sim2.1$ \cite{jones22} -- very similar to the parameters derived for Abell~46.  Just as for Abell~46, the implied surface gravity and temperature lie on a post-RGB evolutionary track for a much lower mass than that derived from the light and radial velocity curve modelling (figure \ref{fig2}).

\subsection{PN~G283.7$-$05.1}

Simultaneous light and radial velocity curve modelling of the eclipsing binary (HW Vir-type) in PN~G283.7$-$05.1 leads to a definitively post-RGB mass determination of $0.34\pm0.05$~M$_\odot$, however the implied luminosity is extremely high for a post-RGB star -- $\log L\sim3.5$ \cite{jones20}.  Similarly, the surface gravity and temperature of the model central star reside in a distinctly post-AGB region of the Kiel diagram (figure \ref{fig2}).  Nonetheless, comparison of the spectra, used to derive the radial velocities of the system, with NLTE models indicates that the surface gravity and temperature from the light and radial velocity curve modelling are indeed consistent with the observed spectroscopy.

\section{Conclusions}

Theoretical studies indicate that post-RGB masses greater than approximately 0.25~M$_\odot$ should lead to observable PNe \cite{hall13}.  Furthermore, comparison of the orbital period distribution of naked post-CE binaries (i.e.\ those that are now too old to reside inside  a visible PNe) -- a significant fraction of which are post-RGB \cite{rebassa-mansergas11} -- and the orbital period distribution of post-CE central stars of PNe indicates that the two are consistent with being drawn from the same population \cite{boffin19}.  As such, there is little reason to believe that the currently known sample of post-CE central stars (now roughly 100) contains a number of post-RGB systems.  However, as yet, no definitively post-RGB central star of a PN has been identified, although a handful of excellent candidates have been found.  If the strongest of the candidates are indeed post-RGB then this reveals important information about the CE -- in particular, given the apparent discrepancy between masses derived by from radial velocities and those based on their loci in the Kiel diagram (Abell~46, Ou~5, PN~G283.7$-$05.1), that post-RGB evolutionary tracks may not be entirely representative of post-CE evolution.  Any difference between post-CE models and reality can provide an invaluable window into the CE process. Additionally, RGB CEs have been speculatively linked to the extreme abundance discrepancies in observed in some PNe. As such, continued attempts to discover and characterise post-RGB central stars may be key not only in constraining the importance of this formation pathway among PNe but also in constraining the CE and its relation to other long-standing astrophysical questions.

%
%
\small  
%
\section*{Acknowledgments}   

D.J. acknowledges support from the Erasmus+ programme of the European Union under grant number 2020-1-CZ01-KA203-078200.

%

%
\end{document}